\documentclass[12pt]{iopart}
\usepackage{graphicx,amsfonts}
\newcommand{\bea}{\begin{eqnarray}}
\newcommand{\eea}{\end{eqnarray}}

\def\p{\partial}
\def\be{\begin{equation}}
\def\ee{\end{equation}}

\def\ra{\rightarrow}
\def\fr{\frac}
\def\a{\alpha}
\def\b{\beta}

\def\d{\delta}
\def\e{\epsilon}

\def\g{\gamma}

\def\l{\lambda}
\def\m{\mu}
\def\n{\nu}
\def\n{\nu}
\def\r{\rho}
\def\s{\sigma}

\def\w{\omega}

\def\d{\delta}

\def\L{\Lambda}

\def\p{\partial}

\def\no{\nonumber}

  \def\cL{{\cal L}}

\begin{document}

\title{Dual Linearised Gravity in Arbitrary Dimensions}
\author{K M Ajith $^{\dag}$, E Harikumar $^{\ddag}$ and M Sivakumar $^{\dag}$}

\address{$^{\dag}$ School of Physics, University of Hyderabad,\\ Central
University P. O, Hyderabad-500046, India\smallskip \\$^{\ddag}$
Instituto de F\'{\i}sica, Universidade de S\~{a}o Paulo,\\ Caixa
Postal 66318, CEP 05315-970, S\~{a}o Paulo - SP, Brazil}
\ead{ph01ph13@uohyd.ernet.in,hari@fma.if.usp.br and
mssp@uohyd.ernet.in}

\begin{abstract}
We construct dual formulation of linearised gravity in first order
tetrad formalism in arbitrary dimensions within the path integral
framework following the standard duality algorithm making use of
the global shift symmetry of the tetrad field. The dual partition
function is in terms of the (mixed symmetric) tensor field
$\Phi_{[\nu_{1}\nu_{2}\cdots\nu_{d-3}]\nu}$ in {\it frame-like}
formulation. We obtain in d-dimensions the dual Lagrangian in a
closed form in terms of field strength of the dual frame-like
field. Next by coupling a source with the (linear) Riemann tensor
in d-dimensions, dual generating functional is obtained. Using
this an operator mapping between (linear) Riemann tensor and
Riemann tensor corresponding to the dual field is derived  and we
also discuss the exchange of equations of motion and Bianchi
identity.
\end{abstract}
 \noindent{\it Keywords}:
Duality, Spin-2 theory, Mixed symmetric
tensor gauge field\\
\maketitle

\section{Introduction}
The construction and study of dual formulation of physical systems
have been of interest always\cite{savit}. The different but
equivalent formulation of a system is useful for studying
different phases or approximation that may not be transparent in
one description. Among the dual formulations in d-dimensions,
duality between scalar field and $d-2$ form theories is
well-known. Currently there are many interesting attempts to
extend this duality to spin-2 theory (linear gravity) in
particular and to arbitrary spin-s theories in general. Part of
the motivation for the former is to understand hidden symmetry in
gravitational theory, if any. An earliest approach to obtain
strong-weak duality for gravitational theory was first made in
\cite{garcia}. An analogue of S duality for 4d linearised gravity
in MacDowell-Mansouri formalism was made in \cite{ nieto}. It has
been shown recently in \cite{casini, hull1,hull,casini1,bek} that
the spin-2 theory can be formulated in terms of a mixed symmetric
tensor having Young symmetry (d-3,1) in d-dimensions. This was
obtained either as on-shell equivalent theories or in a
non-covariant way. Mixed symmetric tensor gauge fields have been
studied, more as a generalisation of $p$-form
theories\cite{DST,curt}. Curtright \cite{curt} in particular has
given a construction in terms of $\phi_{[\m\n]\l}$ with
$\phi_{[[\m\n]\l]}=0~$(here, square bracket [ ] denotes
antisymmetry), which describes linearised gravity in 5-dim.
Zinoviev \cite{zin} has generalised this construction in a first
order formulation for more general cases. Recently, starting from
a first order formulation of spin-2 theory \cite{vasilev},
Boulanger, Cnockaert and Henneaux \cite{henn} have constructed
dual spin-2 theory in a manifestly covariant and off-shell
formulation. These authors have shown that in 4-dimensions, the
dual theory is same as the original Fierz-Pauli (FP)theory while
in 5-dimensions, the dual theory turned out to be the one given by
Curtright\cite{curt}. It has been also shown that in arbitrary
d-dimensions, the dual theory is formulated in terms of field,
$\phi_{[\mu_{1}\mu_{2}\cdots\mu_{d-3}]\nu}$ obeying the condition
of vanishing cyclic sum of indices, i.e.,
$\phi_{[[\mu_{1}\mu_{2}\cdots\mu_{d-3}]\nu]}=0$. This construction
has been also been obtained starting from AdS background by
Matveev and Vasiliev \cite{mv}. The duality invariance of the 4d
linearised gravity action was shown  by Henneaux and Teitelboim
\cite{HT}. There are also attempts to extend dual construction to
non-linear gravity also \cite{Zinovievrecent}. All these studies
have been made in metric-like formulation. In this work we
construct dual theory for s=2 {\it in frame-like formulation}.

Dual theories have been generally obtained based on the
algorithm developed in \cite{bus}. This method relies on the
existence of a global symmetry and has been fruitfully applied in
string theory\cite{alvarez}. This method has also been applied to
show Bosonisation as duality \cite{bosonisation}, to exhibit
different formulations of massive gauge theories as dual
formulations \cite{prd} and also in fractional quantum Hall effect
\cite {bal}(for a review of this method, see \cite{non}). This
raises the question of deriving the dual spin-2 theories applying
this method. In this paper we show that within this standard
duality construction, based on the existence of the global shift
symmetry, dual partition function and the generating functional
for linear gravity in d-dimensions can be obtained in frame-like
formulation

The standard FP theory (which is same as the linear level
approximation of general relativity) has symmetric second rank
tensor $h_{\m\n}$ as the basic field and its Lagrangian is
 \be L
=\frac{1}{2}\partial_{\alpha}h^{\m\n}\p^{\a}
h_{\m\n}-\p_{\n}h^{\n\m}\p^{\b}h_{\b\m}
+\p_{\n}h^{\n\m}\p_{\m}h^{\a}_{~\a}-\fr{1}{2}\p^{\mu}h^{\a}_{~\a}\p_{\m}h^{\b}_{~\b}
\label{FP} \ee which is invariant under $\delta h_{\m\n}=
\p_{\m}\e_{\n}+\p_{\n}\e_{\m}$. Instead of the metric-like
formulation, i.e., symmetric tensor, as the dynamical field, one
can have an equivalent spin-2 description in terms of the frame
field, i.e., tetrad field $e_{\m\n}$ (where no symmetry is assumed
between the indices). The corresponding Lagrangian is \cite{zin}
\be L=\fr{1}{8}T_{[\mu\nu]\lambda}T^{[\mu\nu]\lambda}
+\frac{1}{4}T_{[\mu\nu]\lambda}T^{[\mu\lambda]\nu}-\fr{1}{2}T_{\mu}T^{\mu}
\label{sot} \ee
 where the field strength is $T_{[\a\b]\l}=\p_\a
e_{\b\l}-\p_\b e_{\a\l}$ and $T_{\a}=T_{[\a\b]}^{~~\b}$. Note that
the world index and tangent space index are not distinguished
since we work at the linear level. The above Lagrangian is
trivially  invariant under the transformation \be \delta_{\e}
e_{\m\n}=\p_{\m}\e_{\n} \label{triv} \ee
 and for the specific choice of the coefficients in equation (\ref {sot}) under
 \be
\delta_{\L} e_{\mu\nu} =\Lambda_{[\mu\nu]}. \label{ten} \ee Using
this invariance in equation (\ref{sot}) one can gauge away the
anti-symmetric part of the field $e_{\m\n}$ leaving only the
symmetric part and then equation (\ref{sot}) reduces to Lagrangian
in equation (\ref{FP}). The first order Lagrangian equivalent to
the above one in equation (\ref{sot}) is given by
 \be
L_{I}=-\frac{1}{2}\omega^{\mu[\nu\alpha]}\omega_{\n[\mu\a]}
-\frac{1}{2}\omega^{\m[\nu\a]}T_{[\nu\a]\m}
+\frac{1}{2}\omega_{\mu}\omega^{\mu}-\omega^{\mu}T_{\mu}.
\label{fot} \ee Here $\omega_{\mu}=\w^{\n}_{~\n\m}$. This
Lagrangian is invariant under the transformations in
equation(\ref{triv}) as well as under \be \delta_{\L}\omega_{\mu
[\alpha\beta]}=\partial_{\mu}\Lambda_{[\alpha\beta]},~~~~~
\delta_{\L} e_{\m\n}=\Lambda_{[\m\n]}. \label{gifot} \ee
 Note that this first
order form is valid in d-dimensions and differs from the starting
point of \cite{henn} by a field redefinition. \footnote{The field
redefinition \be
\omega_{\a[\b\gamma]}=Y_{[\b\gamma]\a}+\frac{2}{D-2}\eta_{\alpha [
\beta} Y_{\gamma]\r}^{~~~\r}\ee in (\ref{fot}) leads to the
starting point of \cite{henn} (up to an overall scale).} In this
paper we start with the Lagrangian given in equation (\ref{fot})
instead of FP theory directly and obtain the equivalent dual model
in arbitrary dimensions, but in frame-like formulation.

In the method adopted \cite {bus} one considers a theory which has
a global symmetry and this global symmetry is gauged to a local
one by introducing an appropriate gauge field.  The dual field
strength of the gauge field is then constrained to vanish by means
of a multiplier field. Integrating the multiplier field and the
gauge field (which will become pure gauge now), original theory
results. Instead, integrating out the original field and the gauge
field gives the dual theory where the multiplier field becomes the
dynamical field. It is well known that this method applied to
massless scalar field leads to d-2 form gauge theory. Applied to
Maxwell theory in 4-dimensions, lead again to Maxwell action in
terms of dual vector potential\cite{maxwell}.

The global symmetry which we make use of is shift of the tetrad
field. Hence it is extendable to coupling sources with suitable
fields, such that global symmetry is preserved. Then the
equivalence between the correlation functions can be derived. The
main results obtained here are :
\begin{enumerate}
\item{ Derivation of the dual partition function for linear
gravity, described by the dual  {\it frame-like}
 field,
$\Phi_{[\nu_{1}\nu_{2}\cdots\nu_{d-3}]\nu}$ in a closed form in
terms of field strength.} \item{ By coupling
 source to (linear) Riemann tensor, we
obtain the equivalent dual generating functional and infer the
operator
 mapping between Riemann tensor and its dual equivalent.}
\end{enumerate}

This paper is organized as follows. In section \ref{fotdual} we
consider 4-dim. spin-2 theory described by equation (\ref{fot})
and, derive equivalent theory. In section \ref{5ddual}, deals with
duality in dimensions d$>$4.  In Section \ref{cor} dualisation of
the theory with external source is dealt and the mapping between
the correlators are derived.  Our concluding remarks are given in
section \ref{con}. We use $\epsilon_{0,1,2..d-1}=1$ and the metric
$g_{\mu\nu}={\rm diag}({1,-1,-1,-1,\cdots,-1})$. Appendix gives
the details of duality between the equation of motion and Bianchi
identity.

\section{Dual theory in $3+1$-dimensions }\label{fotdual}

In this section we construct the dual formulation of massless
spin-2 theory described by the Lagrangian in equation (\ref{fot})
in $3+1$ dim. The first-order theory described by the Lagrangian
in equation (\ref{fot}) is invariant under the gauge
transformations given in equation (\ref{gifot}). Apart from this,
the Lagrangian is also invariant under the shift of $e_{\m\n}$
field by a global parameter, i.e., under \be e_{\m\n}\rightarrow
e_{\m\n}+\e_{\m\n}. \ee We gauge this symmetry by introducing a
gauge field $K_{[\m\n]\l}$ which transform as $K_{[\m\n]\l}
\rightarrow K_{[\m\n]\l}+\p_\m\e_{\n\l}-\p_\n\e_{\m\l}$, there by
making the theory invariant under local shift of $e_{\m\n}$ field.
Here the gauge field $K_{[\m\n]\l}$ is a mixed symmetric tensor
having same symmetry in its indices as the field $T_{[\m\n]\l}$.
Now to bring it equivalent to original theory, the dual of the
field strength $\p_{\m}K_{[\n\s]\b}$ should vanish. This is
enforced by a multiplier field $\phi_{\b\r}$ (there is no symmetry
in the indices of multiplier field). Thus the master Lagrangian
which is equivalent to equation (\ref{fot}) is : \bea
L_{M}^{(4)}&=&-\frac{1}{2}\omega^{\mu[\nu\alpha]}\omega_{\n[\mu\a]}+\frac{1}{2}\omega^{\mu}\omega_{\mu}
-\frac{1}{2}\omega^{\beta[\nu\sigma]}(T_{[\nu\sigma]\beta}
-K_{[\nu\sigma]\beta})\no\\
&-&\omega^{\nu}T_{\nu}+\omega^{\nu}K_{\nu}+
\frac{1}{4}\epsilon^{\mu\nu\sigma\r}(\p_{\m}K_{[\n\s]\b})\phi_{\r}^{~\b}.
\label{lagp} \eea This master Lagrangian has to satisfy two
conditions: (1) It must possess all the symmetries of the original
Lagrangian given in equation (\ref{gifot}) and (2) when the
multiplier field equation is imposed, the master Lagrangian should
reduce to the original theory. The first condition is satisfied by
demanding that under the transformation of $e_{\m\n}$ and
$\w_{\n[\a\b]}$ in equation (\ref{gifot}), we also have \be
\delta_{\L}K_{[\n\m]\a}=0,~~~~
\delta_{\L}\phi_{\m\n}=\e_{\m\n\r\s}\L^{[\r\s]} \equiv
\tilde\L_{[\m\n]} \ee It is interesting to note that the
transformation parameter of original field $e_{\m\n}$ and the
multiplier field $\phi_{\m\n}$ are dually related, which means
both undergo transformation by an antisymmetric parameter. The
invariance under the transformation in equation (\ref{triv}) is
trivially present as the fields $e_{\m\n}$ appear in $L_{M}^{(4)}$
only through $T_{[\m\n]\a}$. It is easy to see that the second
condition is satisfied as eliminating $\phi_{\m\n}$ field sets
$K_{[\m\s]\b}=\p_{\m}\e_{\s\b}-\p_{\s}\e_{\n\b}$ which can be used
to redefine $e^{'}_{\m\n}=e_{\m\n}-\e_{\m\n}$ to get back the
original theory. Note that in the action  $L_{M}$, $\phi_{\m\n}$
is arbitrary up to an addition of $\p_\m\tilde\e_\n$. This
translates into gauge symmetry of the dual field.

We start the dualisation from the partition function 
\be 
Z=\int D\phi_{\a\b} De_{\m\n}D\w_{\s[\r\l]} DK_{[\theta\tau]\g}~e^{-i\int d^4 x
L_{M}^{(4)}} 
\ee 
where the Lagrangian is given in equation
(\ref{lagp}). Now to obtain the dual theory we integrate out the
gauge field. As the gauge field appears only linearly in the
action, integrating $K_{[\n\s]\b}$ in the above partition function
results the delta function \be
\d(\fr{1}{2}\omega^{\beta[\nu\sigma]}+\frac{1}{2}(\omega^{\nu}g^{\sigma\beta}-\omega^{\sigma}g^{\nu\beta})+\frac{1}{4}\epsilon^{\mu\nu\sigma\delta}
\p_{\mu}\phi_{\delta}^{~~\beta}) \label {delta} \ee in the
measure. The integration over $\w$ field are now trivial as this
delta function allows one to replace the $\w$ fields in terms of
the multiplier field as given below. \bea \omega^{\nu}&=&
\frac{1}{4}\epsilon^{\nu\mu\g\b}F_{[\mu\g]\beta}
\label{wt}\no\\
\omega^{\beta
[\nu\sigma]}&=&\frac{1}{4}\big[(g^{\beta\sigma}\epsilon^{\nu\mu\g\l}-g^{\beta\nu}\epsilon^{\sigma\mu\g\l})F_{[\mu\g]\lambda}
-2\epsilon^{\nu\sigma\m\g}F_{[\mu\g]}^{~~~\beta})\big] \label{w}
\eea where $F_{[\m\n]\b}=\p_{\m}\phi_{\n\b}-\p_{\n}\phi_{\m\b}$ is
the field strength of multiplier field. It is easy to see that
when these solutions for $\w$ are plugged back in the action, all
$e$ dependent terms vanish and then the integration over $e$ field
just gives a multiplicative infinity which can be absorbed in the
normalization. Thus the partition function corresponding to the
dual theory is \be
 Z=\int D\phi_{\m\n}~e^{-i\int d^4 x {L}_{D}^{(4)}}
\ee
 where the effective Lagrangian is
\be L_{D}^{(4)}=\frac{1}{8}F_{[\mu\g]\lambda}F^{[\mu\g]\lambda}
+\frac{1}{4}F_{[\mu\g]\lambda}F^{[\mu\lambda]\g}
-\frac{1}{2}F^{\mu}F_{\mu} \label{ldual} \ee where
$F_{\m}=F_{\m\n}^{~~~\n}$. Thus we see that the dual description
is in terms of non symmetric, rank$-2$ tensor. It is obvious that
the above dual Lagrangian has $\d\phi_{\m\n}=\p_{\m}\tilde\e_{\n}$
symmetry as $F_{[\mu\g]\lambda}$ has this symmetry.  This
Lagrangian is exactly the same in structure and coefficients as
that in the equation (\ref{sot}). Hence the dual theory also has
the same symmetry $\delta\phi_{\m\nu}=\tilde\L_{[\m\n]}(x).$

This result can also be seen directly by starting from the second
order theory given in equation (\ref{sot}). The corresponding
master Lagrangian is \be L_{M}^{(4)}=\fr{1}{8}\bar T^{[\m\n]\a}
\bar T_{[\m\n]\a}+\fr{1}{4}\bar T^{[\m\n]\a} \bar
T_{[\m\a]\n}-\fr{1}{2}\bar T^{\m}\bar
T_{\m}+\fr{1}{4}\e_{\m\n\a\b}(\p^{\n}K^{[\a\b]\s})\phi^{\m}_{~\s}
\label{sot1} \ee where $ \bar
T^{[\m\n]\a}=T^{[\m\n]\a}-K^{[\m\n]\a}$ and $\bar
T^{\m}=T^{\m}-K^{\m}$. As earlier, we have introduced here the
gauge field $K_{[\m\n]\l}$ elevating the global shift invariance
to a local one.

We start the dualisation from partition function 
\be
Z=\int{D\phi_{\a\b}De_{\m\n}DK_{[\s\r]\l}e^{-i\int{d^{4}xL_{M}^{(4)}}}}
\label{sop} 
\ee 
where the Lagrangian is given in equation
(\ref{sot1}). By varying $K^{[\m\n]\a}$ the equation of motion
following is
 \bea
&&
-\fr{1}{4}T^{[\m\n]\a}+\fr{1}{4}K^{[\m\n]\a}-\fr{1}{4}T^{[\m\a]\n}+\fr{1}{4}T^{[\n\a]\m}+\fr{1}{4}K^{[\m\a]\n}-\fr{1}{4}K^{[\n\a]\m}\nonumber\\&&
+\fr{1}{2}T^{\m}g^{\n\a}-\fr{1}{2}T^{\n}g^{\m\a}-\fr{1}{2}K^{\m}g^{\n\a}+\fr{1}{2}K^{\n}g^{\m\a}-\fr{1}{4}\e^{\s\b\m\n}\p_{\b}\phi_{\s}^{~\a}=0.
\eea
 From this we get
\bea
&\bar T^{\m}=-\fr{1}{4}\e^{\m\b\s\n}F_{[\b\s]\n}\nonumber&\\
& \bar T^{[\m\n]\a}=\fr{1}{4}(g^{\a\m}
\e^{\n\b\s\r}-g^{\a\n}\e^{\m\b\s\r})F_{[\b\s]\r}
-\fr{1}{2}(\e^{\b\s\a\m}F_{[\b\s]}^{~~~\n}-\e^{\b\s\a\n}F_{[\b\s]}^{~~~\m}).&
\label{eqmsot} \eea
  Substituting this back in equation (\ref{sop}), which is equivalent to integrating
out $K^{[\m\n]\a}$ and $e_{\m\n}$ we get the dual
 partition function
 \be
  Z=\int{D\phi_{\m\n}e^{-i\int{d^{4}xL_{D}^{(4)}}}}
 \ee
 where $L_{D}^{(4)}$ is same as given in equation (\ref{ldual}). Thus even directly
from second order theory equation (\ref{sot}) one can arrive at
the dual (second order) theory. The field content and the form of
the dual theory is same as the original second order form equation
(\ref{sot}). This is in agreement with earlier studies. This is
similar to that of Maxwell theory in $3+1$ dimensions. This
completes the dualisation of 4d theory.

It should be noted that the technique adopted here to derive the
dually equivalent formulation is different from Fradkin-Tseytlin
\cite{ft}approach. In contrast to the Fradkin-Tseytlin approach,
the necessary condition in the present approach is the existence
of a global symmetry in the original theory and is not restricted
to a first-order formalism. Moreover,  in this approach, at every
stage gauge invariance of the original theory is preserved.

\section{Dual formulation in dimensions-$d>4$}\label{5ddual}

In this section we first study the dual construction in $5$-dim.
and then generalise it to arbitrary $d$ dimensions. The difference
between $4$-dim. and $5$-dim. case comes in the term involving the
multiplier field. This term, enforcing the flatness condition of
the dual field strength now has $5$-d Levi-Civit\`a symbol and
correspondingly, the multiplier field has a different tensorial
structure compared to that in the $4$-dim. case studied above.

In the present case, thus we start from the master Lagrangian 
\bea
{L}_{M}^{(5)}&=&-\frac{1}{2}\omega^{\mu[\nu\alpha]}\omega_{\n[\mu\a]}
+\frac{1}{2}\omega^{\mu}\omega_{\mu}-
\frac{1}{2}\omega^{\m[\nu\a]}(T_{[\nu\a]\m}-K_{[\nu\a]\m})\nonumber
\\&&-\omega^{\nu}T_{\nu}+\omega^{\nu}K_{\nu}
+
\frac{1}{12}\epsilon^{\mu\n\s\a\g}(\p_{\mu}K_{[\n\s]\b})\phi_{\a\g}^{~~~\b}.
\label{5d1} 
\eea 
Here the field strength is defined as
$F_{[\mu\tau\gamma]\lambda}
=\partial_{[\mu}\phi_{[\tau\gamma]]\lambda}$ and
$\phi_{[\tau\gamma]\beta}$ field is mixed symmetric Lagrange
multiplier field. As in 4-dim case, demanding $L_{M}^{(5)}$ to
have the original spin-2 symmetry requires
$\delta_{\Lambda}K_{[\n\s]\b}=0$ and
$\d_{\L}\phi_{[\m\n]\l}=\frac{3}{2!}\e_{\m\n\l\r\s}\L^{[\r\s]}(x)
=\tilde\L_{[\m\n\l]} $. The corresponding partition function is
\be
 Z=\int
D\phi_{[\a\b]\l} De_{\m\n}D\w_{\s[\r\l]}DK_{[\theta\tau]\lambda}~e^{-i\int
d^5 x {L}_{M}^{(5)}}. 
\ee
 The partition function of the dual
theory is obtained by integrating over the gauge field
$K_{[\m\n]\l}$, original fields $e_{\m\n}$ and $\w_{\m[\n\l]}$ as
in the $4$-dim. The gauge transformation of the dual field, which
is the multiplier field $\phi_{\a\g}^{~~~\b}$, results from the
arbitrariness in its definition in the master Lagrangian.

 $ {L}_{M}^{(5)}$ equation(\ref{5d1})is invariant under the following
 transformation(up to a total derivative)
 \be
\phi_{\a\b,\gamma}\rightarrow\phi_{\a\b,\gamma}
+\partial_{\a}\tilde\e_{\b\gamma}-\partial_{\b}\tilde\e_{\a\gamma}
\ee where $\tilde\e_{\b\gamma}$ is an arbitrary second rank
tensor. The fields strength $F_{[\mu\tau\gamma]\lambda}$ is
invariant under this transformation.

The $K_{[\m\n]\l}$ integrations leads to the delta function
condition (as in the $4$-dim) which now involve third rank mixed
symmetric field and the $\w_{[\m\n]\l}$ field and is given by 
\be
\delta(\w^{\b[\n\s]}+(\w^{\n}g^{\s\b}-\w^{\s}g^{\n\b})+\fr{1}{6}\e^{\m\a\g\n\s}F_{[\m\a\g]}^{~~~~\b}).
\ee 
The partition function of the dual theory is obtained by
integrating over $\w_{\m[\n\l]}$ which can be done similar to 4d
case leading to \be
 Z=\int D\phi_{[\a\b]\l}~e^{-i\int d^5 x{L}_{D}^{(5)}}
 \ee
where \be
L_{D}^{(5)}=-\frac{1}{6}F_{[\mu\nu\gamma]\lambda}F^{[\mu\nu\gamma]\lambda}-\frac{1}{4}F_{[\mu\nu\gamma]\lambda}F^{[\mu\nu\lambda]\gamma}+\frac{3}{4}F_{[\mu\gamma\r]}^{~~~~~\r}F^{[\mu\gamma\nu]}_{~~~~~~\nu}.
\label{5d} \ee The gauge symmetries of the dual Lagrangian are
given by \bea
\delta_{\tilde\e}\phi_{\a\b,\gamma}&=&\partial_{\a}\tilde\e_{\b\gamma}-\partial_{\b}\tilde\e_{\a\gamma}
\label{gt0}\\
\delta_{\Lambda}\phi_{\a\b,\gamma}&=&\tilde\Lambda_{[\a\b\gamma]}.
\label{gt} \eea

These are the generalisation of equation(\ref{triv}) and equation
(\ref{ten}) to 5d in dual frame-like formulation. This is the same
as the Lagrangian for spin-$2$ theory constructed in \cite{zin}
using the $(2,1)$ tensor gauge field $\phi_{[\delta\gamma]\beta}$.

In \cite{henn} dual form of FP theory in 5-dim. was shown to
described by the Lagrangian \be
\cL=-\frac{1}{6}(F_{[\m\n\l]\g}F^{[\m\n\l]\g}-3F_{[\m\n\l]}^{~~~~~\l}F^{[\m\n\r]}_{~~~~~\r}).
\label{curt} \ee Here also
$F_{[\m\n\l]\g}=\partial_{[\m}\phi_{\n\l]\g}$ but the fields obey
the condition $\phi_{[[\m\n]\l]}=0$
 which is absent in  the dual
theory derived in equation (\ref{5d}). Hence the question arises
as to how equation (\ref{5d}) is related to equation (\ref{curt})
as both are dual to FP theory in $5$ dim.

The Lagrangian in equation (\ref{curt}) is in metric-like
formulation as the field satisfies $\phi_{[[\m\n]\l]}=0$, unlike
the one given in equation (\ref {5d}) which is in frame-like
formulation. The connection between them can be seen by working in
the gauge $\phi_{[[\m\n]\l]}=0$, using equation (\ref{gt}). Then
the gauge transformation  in the equation (\ref{gt0}), which
preserves this gauge, is

\be
\delta\phi_{[\n\gamma]\lambda}=\frac{1}{3}[\partial_{\n}A_{[\gamma\lambda]}
+\partial_{\gamma}A_{[\lambda\n]}-2\partial_{\lambda}A_{[\gamma\n]}]
+ \p_{[\n}S_{\gamma]\lambda} \label{gicur} \ee where $A_{\a\b}=-
A_{\b\a}$ and $ S_{\a\b}=S_{\b\a}$.  Note that the most general
Lagrangian involving $F_{[\mu\nu\l]\r}$ is \be
L=-\big(F_{[\mu\n\gamma]\lambda}F^{[\mu\n\gamma]\lambda}+B
F_{[\mu\nu\g]\lambda}F^{[\mu\nu\l]\gamma}
+CF_{\mu\gamma\l}^{~~~~\l}F^{\mu\gamma\nu}_{~~~~\nu}\big).
\label{mosyg} \ee

$F_{[\m\n\gamma]\l}$ is invariant under $S_{\a\b}$ and by demanding
the symmetry of (\ref{mosyg}) under $A_{[\gamma\lambda]}$ given in
equation (\ref{gicur})  gives the condition $3+B+C=0$. Thus there
is a one-parameter of theories and  for a choice of B=0, C takes
-3 which is Curtright's Lagrangian.

Now extending this procedure to arbitrary dimensions is straight
forward. The starting Lagrangian is 
\bea
L_{M}^{(d)}&=&-\frac{1}{2}\omega^{\mu[\nu\alpha]}\omega_{\n[\mu\a]}+\frac{1}{2}\omega^{\mu}\omega_{\mu}
-\frac{1}{2}\omega^{\m[\nu\a]}(T_{[\nu\a]\m}-K_{[\nu\a]\m})\nonumber\\
&-&\omega^{\nu}T_{\nu} +\omega^{\nu}K_{\nu}
+\frac{1}{2(d-2)!}\epsilon^{\mu_{1}\mu_{2}\mu_{3}\mu_{4}\cdots\mu_{d}}\partial_{\mu_{1}}K_{[\mu_{2}\mu_{3}]\beta}\phi_{[\mu_{4}\cdots\mu_{d}]}^{~~~~~~~~\beta}.
\label{need} 
\eea 
Here also in order to maintain the original
spin-2 symmetry, the gauge field and multiplier field undergoes
the corresponding transformations,
$\delta_{\Lambda}K_{[\mu_{2}\mu_{3}]\b}=0$ and
$\d_{\L}\phi_{[\m_{1}\m_{2}\cdots\m_{d-3}]\n}=\frac{(d-2)}{2!}\e_{\m_{1}\m_{2}\cdots\m_{d-3}\nu\s\l}\L^{\s\l}(x)
= \tilde\Lambda_{[\m_{1},\m_{2}\cdots\m_{d-3}\n]}.$

The gauge freedom associated with the dual field ( i.e, multiplier
field) owes to the arbitrariness in its definition in the action
for $L_{M}^{(d)}$ by \bea
\delta_{\tilde\e}\phi_{[\m_{1}\m_{2}\cdots\m_{d-3}]\n}=
\p_{[\m_{1}}\tilde\e_{\m_{2}\cdots\m_{d-3}]\n} \label{dgt}\eea
where $\tilde \e_{\m_{1}\cdots\m_{d-4}\n}$ is antisymmetric in
first $\mu_{1}$to $\mu_{d-4}$ indices.

Now after integrating out the gauge field
$K_{[\mu_{1}\mu_{2}]\beta}$ and the original fields $e_{\mu\nu} ,
\w_{\m[\n\l]}$ as in 4-dim. and 5-dim. we get the dual Lagrangian
in frame-like field as \bea
L_{D}&=&\frac{1}{2(d-2)(d-2)!^{2}}\epsilon^{\mu_{1}\mu_{2}\mu_{3}\m_{4}\cdots\mu_{d}}\epsilon_{\nu_{1}\mu_{2}\nu_{3}\n_{4}\cdots\nu_{d}}F_{[\mu_{1}\mu_{4}\cdots\mu_{d}]\mu_{3}}F^{[\nu_{1}\nu_{4}\cdots\nu_{d}]\nu_{3}}\nonumber\\&&
-\frac{1}{2(d-2)!^{2}}\epsilon^{\mu_{1}\mu_{2}\mu_{3}\mu_{4}\cdots\mu_{d}}F_{[\mu_{1}\mu_{4}\cdots\mu_{d}]}^{~~~~~~~~~~\n_{2}}\epsilon_{\nu_{1}\nu_{2}\mu_{3}\nu_{4}\cdots\nu_{d}}F^{[\nu_{1}\nu_{4}\cdots\nu_{d}]}_{~~~~~~~~~~\mu_{2}}
\label{d} \eea where the field strength
$F_{\mu_{1}\cdots\mu_{d-2},\n}$ is
$\partial_{[\mu_{1}}\phi_{\mu_{2}\cdots\mu_{d-2}]\n}$~and is
invariant under~$\tilde\e$~transformation in equation (\ref{dgt}).
The symmetries of this dual Lagrangian are \bea
\delta_{\tilde\e}\phi_{[\m_{1}\m_{2}\cdots\m_{d-3}]\n}&=
&\p_{[\m_{1}}\tilde\e_{\m_{2}\cdots\m_{d-3}]\n}\nonumber\\
\delta_{\Lambda}\phi_{[\m_{1}\m_{2}\cdots\m_{d-3}]\n}&=&
\tilde\L_{[\m_{1}\m_{2}\cdots\m_{d-3}\n]} \label{gtd} \eea

This agrees with equation (\ref{ldual}) and equation (\ref{5d})
for $d=4$ and $d=5$ respectively. Also note that for $d=6$, the
dual Lagrangian obtained here is same as the Lagrangian for
$\phi_{[\mu_{1}\mu_{2}\mu_{3}]\mu_{4}}$ field constructed in
\cite{zin}. Thus we get the general Lagrangian extending Zinoviev
construction to d-dimensions and its identification as dual spin-2
theory follows from equation (\ref{d}).
 The above Lagrangian can be written (apart from an
overall factor) in a more convenient form as
 \bea
\fl L_{D}= (-1)^{d} \Big[F_{[\mu_{1}\mu_{2}\cdots
\mu_{d-2}]\n}F^{[\mu_{1}\mu_{2}\cdots\mu_{d-2}]\n}+\frac{(d-2)}{(d-3)}F_{[\mu_{1}\mu_{2}\cdots\mu_{d-2}]\n}F^{[\mu_{1}\mu_{2}\cdots\mu_{d-3}\n]\mu_{d-2}}\nonumber\\
-\frac{(d-2)^{2}}{(d-3)}F_{[\mu_{1}\mu_{2}\cdots\mu_{d-3}\a]}^{~~~~~~~~~~~~~~\a}F^{[\mu_{1}\mu_{2}\cdots\mu_{d-3}\r]}_{~~~~~~~~~~~~~~\r}\Big].
\label{df} \eea This form of the dual Lagrangian in a {\it
frame-like} formulation is an important result of this work.

This form of Lagrangian can be argued from the gauge invariance of
the dual theory. The most general Lagrangian involving the field
strength $F_{[\m_{1}\m_{2} \cdots \m_{d-2}]\n }$(which is
invariant under $\tilde\e$ transformation given in equation
(\ref{gtd})) is of the form \bea \fl L_{D}= (-1)^{d}
\Big[F_{[\mu_{1}\mu_{2}\cdots
\mu_{d-2}]\n}F^{[\mu_{1}\mu_{2}\cdots\mu_{d-2}]\n}+B
F_{[\mu_{1}\mu_{2}\cdots\mu_{d-2}]\n}F^{[\mu_{1}\mu_{2}\cdots\mu_{d-3}\n]\mu_{d-2}}\nonumber\\
+C
F_{[\mu_{1}\mu_{2}\cdots\mu_{d-3}\a]}^{~~~~~~~~~~~~~~\a}F_{[\mu_{1}\mu_{2}\cdots\mu_{d-3}\r]}^{~~~~~~~~~~~~~~\r}\Big]
\label{mgld} \eea and by demanding the invariance given in
equation $(\ref{gtd})$ under $\tilde\Lambda$, the parameters  get
fixed as \be B=\fr{(d-2)}{d-3},~~~~ C=-\fr{(d-2)^{2}}{(d-3)}.
\label {param} \ee

A metric-like formulation of Curtright's theory has been
formulated in \cite{aulakh}. The connection of our frame-like
formulation to the metric-like formulation can be obtained by
fixing the gauge $\phi_{[[\m_{1}\m_{2}\cdots\m_{d-3}]\n]}=0$ using
$\tilde\L$ transformation in (\ref{gtd}).  Then the gauge
transformation, which preserves this gauge under $\tilde \e $
becomes

\bea
\fl\delta\phi_{[\m_{1}\m_{2}\cdots\m_{d-3}]\n}=\p_{\m_{1}}A_{\m_{2}\cdots\m_{d-3}\n}+\p_{\m_{2}}A_{\m_{3}\cdots\m_{d-3}\n
\m_{1}}+\cdots-(d-3)~\p_{\n}A_{\mu_{1}\mu_{2}\cdots\m_{d-3}}+
\nonumber\\
\partial_{[\mu_{1}}\widehat{e}_{\mu_{2}\cdots\m_{d-3}]\n}
\label{gcu} \eea where A is totally antisymmetric tensor and
$\widehat{e}_{[\mu_{2}\cdots\m_{d-3}]\n}$is a mixed symmetric
tensor with totally antisymmetric part removed. The field strength
is invariant under $\widehat{e}$ parameter and invariance of
equation (\ref{mgld}) under A parameter gives a relation between
the coefficients as $(d-2)+B+C=0$. Thus there is a one-parameter
family of theories in metric-like formulation.

\section{Dual generating functional in $d$ $\ge$ 4}\label{cor}

In this section we obtain the equivalence between the generating
functionals and derive the mapping between the observables of the
equivalent theories. For this we couple the original field to an
appropriate source and it is natural to couple $e_{\m\n}J^{\m\n}.$
But this does not have the shift symmetry of the field. The source
has to couple to a combination of field which has the global shift
symmetry and also all the other gauge symmetries of the original
Lagrangian. Since we wish to use first order Lagrangian, the
relevant gauge transformations are that given in equation
(\ref{gifot}) and we couple the gauge invariant observable
$\partial_{\mu}\omega_{\nu[\alpha\beta]}-\partial_{\nu}\omega_{\mu[\alpha\beta]}
$ to the source $J_{[\m\n][\a\b]}$. The starting point now is the
Lagrangian \be L_{J}^{(4)}=
L_{I}+J_{[\m\nu][\alpha\beta]}(\partial_{\mu}\omega_{\nu[\alpha\beta]}-\partial_{\nu}\omega_{\mu[\alpha\beta]})
\label{withs} \ee where $L_{I}$ is equation (\ref{fot}). The gauge
invariant observable chosen is linear Riemann curvature tensor.
First we illustrate the operator correspondence in 4d case. Now
note that global shift symmetry in equation(\ref{fot}) not
modified. Hence the master Lagrangian with the present
augmentation of source $J_{[\m\n][\alpha\beta]}$ coupling is \bea
L_{M}^{(4)}(J)&=&L_{M}^{(4)}+J_{[\mu\nu][\alpha\beta]}(\partial^{\mu}\omega^{\nu[\alpha\beta]}-\partial^{\nu}\omega^{\mu[\alpha\beta]})
\label{j4d} \eea where $L_{M}^{(4)}$ is given in equation
(\ref{lagp}). Integrating over K field from the partition function
corresponding to the above Lagrangian $L_{M}^{(4)}(J)$ gives the
same delta function given in ($\ref{delta}$). Integrating over
fields $\w_{\m[\n\l]}$ and $e_{\m\n}$ result in the (dual)
partition function corresponding to the (dual) Lagrangian \bea
L_{D}^{(4)}[J]&=&L_{D}^{(4)}
-\frac{1}{2}J_{[\mu\nu][\alpha\beta]}\epsilon^{\alpha\beta\r\lambda}(\partial^{\mu}\partial_{\rho}\phi_{\lambda}^{~~\nu}-\partial^{\nu}\partial_{\rho}\phi_{\lambda}^{~~\mu})\nonumber\\
&+&\frac{1}{4}J_{[\mu\nu][\alpha\beta]}\big\{\partial^{\mu}(g^{\nu\beta}\epsilon^{\alpha\r\lambda\g}\partial_{\rho}\phi_{\lambda\gamma}-g^{\n\alpha}\epsilon^{\beta\r\l\g}\partial_{\rho}\phi_{\lambda\gamma})\nonumber\\
&-&\partial^{\nu}(g^{\mu\b}\epsilon^{\alpha\r\l\gamma}\partial_{\rho}\phi_{\lambda\gamma}-g^{\mu\alpha}\epsilon^{\b\rho\l\gamma}\partial_{\rho}\phi_{\lambda\gamma})\big\}
\label{so4df} \eea where $L_{D}^{(4)}$ is given in equation
(\ref{ldual}). Note here that the terms coupled with source in
above equation (\ref{so4df}) has
$\delta\phi_{\m\n}=\tilde\Lambda_{[\m\n]}(x)$ symmetry. Using this
symmetry, the antisymmetric part of $\phi_{\mu\nu}$ can be gauge
fixed to zero retaining only the symmetric part $\phi^{s}_{\m\n}$.
Then the terms in the second and third line of equation
(\ref{so4df}) involving trace of $J_{[\m\n] [\a\b]}$ vanish.
Non-vanishing source coupling terms is
$-\frac{1}{2}J_{[\mu\nu][\alpha\beta]}\epsilon^{\rho\alpha\beta\lambda}(\partial^{\mu}\partial_{\rho}\phi_{\lambda}^{~~\nu}-\partial^{\nu}\partial_{\rho}\phi_{\lambda}^{~~\mu})\equiv-J{\tilde
R}_{(L)}(\Phi^{s})$ where $ {\tilde R}_{(L)}(\Phi^{s})$ is dual
linear Riemann tensor. Usually in the mapping between the n-point
correlators, dually equivalent n-point functions gets
contributions from contact terms. Here similar terms can be seen
to arise if in the partition function of the starting first-order
theory equation (\ref {fot}), the $\w_{\m[\a\b]}$ field integrals
are done. The resulting second-order theory is of the form \bea
L(J)&=&L-J_{[\m\n][\a\b]}\p^{\n}(T^{[\m\a]\b}-T^{[\m\b]\a}-T^{[\a\b]\m})\nonumber\\&&-30\partial_{\n}J^{[\m\n][\a\b]}\p^{\r}J_{[\m\r][\a\b]}
+72\p^{\n}J_{[\a\n][\b\m]}g^{\a\b}\p_{\r}J^{[\s\r][\g\m]}g_{\s\g}
\eea where $L$ is given in equation (\ref{sot}). Notice that the
term linearly coupling to $J_{[\m\n][\a\b]}$ can be identified as
the Riemann tensor when we use the invariance of $e_{\m\n}$ field
under equation (\ref{ten}). Apart from contact term, by
functionally differentiating the partition functions corresponding
to the Lagrangians in equation (\ref{j4d}) and equation
(\ref{so4df})with the source, we get \be <R_{\a\b\m\n}
(x)~R_{\s\r\g\d}(y)> = <\tilde R_{\a\b\m\n}(x)~ \tilde
R_{\s\r\g\d}(y)>_{dual} \ee where $\tilde
R_{\a\b\m\n}=\e_{\a\b\r\s}R_{\r\s\m\n}$.\\ 
Thus there is a
duality mapping
$R_{\a\b\m\n}(h)\ra\e_{\m\n\r\s}R^{\r\s,\a\b}(\Phi^{s})$. Next we
obtain the map in 5-dim.  Proceeding in the same way one will have
the dual Lagrangian with source coupled \bea
L_{D}^{(5)}(J)&=&L_{D}^{(5)}-
\frac{1}{18}J_{[\mu\nu][\alpha\beta]}\big\{\partial^{\mu}(g^{\nu\beta}\epsilon^{\alpha\r\tau\gamma\l}F_{[\rho\tau\gamma]\lambda}-g^{\nu\alpha}\epsilon^{\beta\r\tau\gamma\l}F_{[\rho\tau\gamma]\lambda})\nonumber\\&
-&\partial^{\nu}(g^{\mu\beta}\epsilon^{\alpha\r\tau\gamma\l}F_{[\rho\tau\gamma]\lambda}-g^{\mu\a}\epsilon^{\beta\r\lambda\tau\gamma\l}F_{[\rho\tau\gamma]\lambda})\big\}\no\\
&-&\frac{1}{6}J_{[\mu\nu][\alpha\beta]}\epsilon^{\alpha\beta\r\tau\gamma}(\partial^{\mu}F_{[\rho\tau\gamma]}^{~~~~\nu}-\partial^{\nu}F_{[\rho\tau\gamma]}^{~~~\mu})
\label{5dsf} \eea where $L_{D}^{(5)}$ is given in equation
(\ref{5d}). Now this $L_{D}^{(5)}$ also has
$\delta\phi_{[\mu_{1}\mu_{2}]\mu_{3}}=\tilde\L_{[\mu_{1}\m_{2}\mu_{3}]}(x)$
invariance and using this (as in 4-d) terms involving trace of
$J_{[\m\n][\a\b]}$ can be set to vanish. From this we can identify
the (linear) Riemann tensor as a function of dual field as
$(\partial^{\mu}F_{\rho\tau\gamma}^{~~~\nu}-\partial^{\nu}F_{\rho\tau\gamma}^{~~~\mu})$.
As in the earlier case, we get the mapping $R_{\m\n,\a\b} \ra
\e_{\a\b\rho\tau\gamma}
(\p_{\mu}F^{[\rho\tau\gamma]}_{~~~~~\n}-\p_{\nu}F^{[\rho\tau\gamma]}_{~~~~~\mu})$.
This is easily generalised to d-dimensions where the dual
Lagrangian with source coupling is \bea \fl
L_{D}^{(d)}(J)=L_{D}^{(d)}
-\frac{1}{(d-2)!}J_{[\mu\nu][\alpha\beta]}\epsilon^{\mu_{1}\alpha\beta\mu_{4}\cdots\mu_{d}}(\partial^{\mu}F_{[\mu_{1}\mu_{4}\cdots\mu_{d}]}^{~~~~~~~~~~\nu}-\partial^{\nu}F_{[\mu_{1}\mu_{4}\cdots\mu_{d}]}^{~~~~~~~~~~\mu})\no\\
\fl+\frac{J_{[\mu\nu][\alpha\beta]}}{(d-2)(d-2)!}\big\{\partial^{\mu}(\epsilon^{\mu_{1}\alpha\mu_{3}\mu_{4}\cdots\mu_{d}}F_{[\mu_{1}\mu_{4}\cdots\mu_{d}]\mu_{3}}g^{\beta\nu}-\epsilon^{\mu_{1}\beta\mu_{3}\mu_{4}\cdots\mu_{d}}F_{[\mu_{1}\mu_{4}\cdots\mu_{d}]\mu_{3}}g^{\alpha\nu})\nonumber\\
-\partial^{\nu}(\epsilon^{\mu_{1}\alpha\m_{3}\mu_{4}\cdots\mu_{d}}F_{[\mu_{1}\mu_{4}\cdots\mu_{d}]\m_{3}}g^{\beta\mu}-\epsilon^{\mu_{1}\beta\m_{3}\mu_{4}\cdots\mu_{d}}F_{[\mu_{1}\mu_{4}\cdots\mu_{d}]\mu_{3}}g^{\alpha\mu})\big\}
\eea where $L_{D}^{(d)}$ is given in equation (\ref{d}). Here we
have identified
$~\frac{1}{(d-2)!}(\partial^{\mu}F_{[\mu_{1}\mu_{4}\cdots\mu_{d}]}^{~~~~~~~~~~\nu}-\partial^{\nu}F_{[\mu_{1}\mu_{4}\cdots\mu_{d}]}^{~~~~~~~~~~\mu})~$
as the dual (linearised) Riemann tensor in terms of field
strength. Thus we obtain a dual formulation of generating
functional with source coupled to linear Riemann tensor, providing
a mapping between Riemann tensor and dual Riemann tensor \be
R^{\m\n\a\b}\ra\epsilon^{\alpha\beta\m_{1}\mu_{4}\cdots\mu_{d}}{\cal
R}_{\mu_{1}\mu_{4}\cdots\mu_{d}}^{~~~~~~~~~\nu\mu} \label{dm} \ee
where the Riemann curvature tensor for the dual field
$\Phi_{\m_{4}\m_{5}...\m_{d}}^{~~~~~~~~~\n}$ is \be {\cal
R}_{\mu_{1}\mu_{4}\cdots\mu_{d}}^{~~~~~~~~~\nu\mu}=
(\partial^{\mu}
F_{\mu_{1}\mu_{4}\cdots\mu_{d}}^{~~~~~~~~~\nu}-\partial^{\nu}F_{\mu_{1}\mu_{4}\cdots\mu_{d}}^{~~~~~~~~~\mu}).
\label{dm1} \ee

\section{Conclusion}\label{con}

In this paper we have constructed the dual linearised gravity
theory in arbitrary dimensions in {\it frame-like} formulation. We
start with a frame-like formulation of linearised gravity, in a
form equivalent to Fierz-Pauli theory and obtain the dual theory
using the global shift of tetrad field. Using this method we have
obtained the partition function in frame like formulation for dual
theory in arbitrary dimensions.  The dual theory obtained in
$5$-dim. and $6$-dim coincide with the theory considered by
Zinoviev for mixed symmetric tensor gauge fields $\phi_{[\m\n]\l}$
and $\phi_{[\m\n\l]\s}$ respectively. We have given the dual
theory in arbitrary dimensions in a closed form (up to an over all
factor) as a sum of three terms. The relative coefficients are
crucial for the Lagrangian to have the required invariance of dual
spin-2 theory. Next we extend this to provide equivalence at the
level of n-point correlators, by coupling source to gauge
invariant observable, the (linear) Riemann tensor. This gives
Riemann tensor in terms of the dual gravitational field in
d-dimensions. We thus show that the dual s=2 theory can be
obtained at the level of partition function and generating
functional within the standard dual procedure.

We can also see that duality mapping derived here between the
Riemann tensor for the $h_{\m\n}$ field and that of the dual field
implies the duality in the sense of exchanging Bianchi identity
and equation of motion. The details are in the appendix.

Although we have considered the dual formulation only at quadratic
level (non-interacting), it is desirable to extend this to cubic
and higher order (interacting) terms. There are several arguments
indicating that duality symmetry must exist at interacting case
also. Hence the free-field analysis carried out here in a
constructive scheme should be considered as a first step towards
the more fundamental problem of constructing dual gravity
including non-linearities. But in recent times several no-go
theorems has been proved \cite{nogo}, showing dual formulation of
non-linear gravity should fall beyond the scope of conventional
perturbative local field theory.

\ack {EH thanks FAPESP for support through grant 03/09044-9. MS
thanks Associateship Scheme for visit to The Abdus Salam ICTP,
where part of the work was done and acknowledges discussion with
L. Alvarez-Gaume. MS acknowledges DST (India)for support through a
project. We thank the referees for several useful suggestions.}

\appendix

\section{ Duality between the equation of motion and Bianchi Identity}

In this appendix we show that duality mapping derived between
curvature tensor in terms of $ h_{\m\n}$ and the dual field
implies duality exchange between equation of motion and Bianchi
identity. Using the duality map in equation (\ref{dm}) we first
re-express the Ricci tensor $R_{\mu\nu}$ in terms of the dual
field as \bea
R_{\mu\nu}&=&\epsilon_{\alpha\nu\rho_{3}\dots\rho_{d}}
\Big(\partial^{\alpha}F^{[\rho_{3}\dots\rho_{d}]}_{~~~~~~~\mu}-\partial_{\mu}F^{[\rho_{3}\dots\rho_{d}]\alpha}\Big)\no\\
&=&-\epsilon_{\alpha\nu\rho_{3}\dots\rho_{d}} {\cal
R}^{[\rho_{3}\dots\rho_{d}]\a}_{~~~~~~~~~\mu}. \label{eleqn} \eea
Note here that the Ricci tensor of $h_{\m\n}$ is thus mapped to
the Bianchi identity of the Riemann curvature tensor corresponding
to the dual field. Using this, we see that the equation of motion
of (sourceless) linear gravity (in terms of $h_{\mu\nu}$ field)
becomes \be
R_{\mu\nu}=0\longrightarrow\epsilon_{\alpha\nu\rho_{3}\dots\rho_{d}}
{\cal R}^{\rho_{3}\dots\rho_{d}\a}_{~~~~~~~~\mu}=0 \ee showing
that the equation of motion for $h_{\m\n}$ gets mapped to the
Bianchi identity for the dual curvature under the map derived in
equation (\ref{dm}). Under this duality map in equation
(\ref{dm}), the Bianchi identity $R_{[\alpha\beta\mu]\nu}=0$
becomes 
\bea
R_{[\alpha\beta\mu]\nu}=0&\longrightarrow&\epsilon_{\alpha\beta\mu\sigma_{4}\dots\sigma_{d}}
\epsilon^{\mu\nu\rho_{3}\dots\rho_{d}}
(\partial^{\alpha}F_{\rho_{3}\dots\rho_{d}}^{~~~~~\beta}-\partial^{\beta}F_{\rho_{3}\dots\rho_{d}}^{~~~~~\alpha})\no\\
&=& \epsilon_{\alpha\beta\mu\sigma_{4}\dots\sigma_{d}}
\epsilon^{\mu\nu\rho_{3}\dots\rho_{d}} {\cal
R}_{\rho_{3}\dots\rho_{d}}^{~~~~~~~\b\a}=0. \label{do2} 
\eea 
Now
the equation of motion following from the Lagrangian in equation
(\ref{df}) is
\begin{equation}
\fl 2\partial^{\mu_{1}}F_{[\mu_{1}\mu_{2}\dots\mu_{d-2}]\nu}
+\frac{1}{(d-3)}\partial^{\mu_{1}}F_{[\mu_{1}\mu_{2}\dots\mu_{d-3}|\n|\mu_{d-2}]}-\frac{(d-2)}{(d-3)}~\partial^{\mu_{1}}F_{[\mu_{1}\mu_{2}\dots\mu_{d-3}|\alpha}^{~~~~~~~~~~~~~\alpha|}g_{\mu_{d-2}]\nu}=0
\label{do1}
\end{equation}
Since the dual map in equation (\ref{dm}) is derived, only after
gauge symmetry
$\delta\phi_{\Lambda}=\tilde\Lambda_{\mu_{1}\dots\mu_{d-3},\nu}$
is fixed, we have to do the same for the above equation of motion
also. With this the equation (\ref{do1}) becomes
 \be
-\partial^{\mu_{1}}F_{[\mu_{1}\mu_{2}\dots\mu_{d-2}]\nu}+\partial^{\mu_{1}}F_{[\mu_{1}\mu_{2}\dots\mu_{d-3}|\alpha}^{~~~~~~~~~~~~~\alpha|}g_{\mu_{d-2}]\nu}=0
\label{do3} \ee which is identical to the right hand side of
equation (\ref{do2}). Thus we see that the Bianchi identity of
linear gravity goes to the equation of motion under the duality
map.

Using  the duality map in equation (\ref{dm}) the other Bianchi
identity $\partial_{[\mu}R_{\nu\rho]\alpha\beta}=0$ becomes 
\be
\epsilon_{\mu\nu\rho\sigma_{4}\dots\sigma_{d}}\partial^{\mu}(\epsilon^{\alpha\beta\rho_{3}\dots\rho_{d}}[\partial^{\nu}F_{\rho_{3}\dots\rho_{d}}^{~~~~~~\rho}-
\nu\leftrightarrow\rho])=0 
\ee 
which is satisfied identically
showing that the dual map is consistent with the Bianchi identity.

\end{document}